\documentclass[fleqn,twoside]{article}
\usepackage{espcrc2}
\usepackage{epsfig}
\usepackage{xspace}

\usepackage{graphicx}
\usepackage[figuresright]{rotating}

\newcommand{\AmS}{{\protect\the\textfont2
\newcommand{\kg}{\ensuremath{\mbox{kg}}\xspace}
\newcommand{\GeV}{\ensuremath{\mbox{GeV}}\xspace}
\newcommand{\MeV}{\ensuremath{\mbox{MeV}}\xspace}
\newcommand{\GeVc}{\ensuremath{\mbox{GeV}/c}\xspace}
\newcommand{\MeVc}{\ensuremath{\mbox{MeV}/c}\xspace}
\newcommand{\T}{\ensuremath{\mbox{T}}\xspace}
\newcommand{\cmsq}{\ensuremath{\mbox{cm}^2}\xspace}
\newcommand{\msq}{\ensuremath{\mbox{m}^2}\xspace}
\newcommand{\cm}{\ensuremath{\mbox{cm}}\xspace}
\newcommand{\mm}{\ensuremath{\mbox{mm}}\xspace}
\newcommand{\micron}{\ensuremath{\mu \mbox{m}}\xspace}
\newcommand{\ns}{\ensuremath{\mbox{ns}}\xspace}
\newcommand{\m}{\ensuremath{\mbox{m}}\xspace}
\newcommand{\s}{\ensuremath{\mbox{s}}\xspace}
\newcommand{\ms}{\ensuremath{\mbox{ms}}\xspace}
\newcommand{\mrad}{\ensuremath{\mbox{mrad}}\xspace}

  A\kern-.1667em\lower.5ex\hbox{M}\kern-.125emS}}

\def\Journal#1#2#3#4{{#1} {\bf #2}, #3 (#4)}

\def\NIMA{{\em Nucl. Instrum. Methods} A}

\def\PLB{{\em Phys. Lett.}  B}
\def\PRL{\em Phys. Rev. Lett.}

\def\be{\begin{equation}}
\def\ee{\end{equation}}
\def\bea{\begin{eqnarray}}
\def\eea{\end{eqnarray}}

\title{Initial results from the HARP experiment at CERN}

\author{J.J. Gomez-Cadenas
\address[MCSD]{IFIC and  Departamento de Fisica, At\'omica y Nuclear,\\ 
P.O. Box 22085, E-46071 Valencia, Spain}
\thanks{This work has been supported in part by research grants from Spanish
Ministry of Science and Generalitat Valenciana}
}

\begin{document}

\begin{abstract}
Initial results on  particle yields obtained by the
HARP experiment are presented. The measurements correspond to proton--nucleus
collisions at beam energies of 12.9 $~GeV/c$
and for a thin Al target of  $5\%$ interacion legth.
The angular range considered is between 10 and 250 $~mrad$. 
This results are the first step
in the upcoming measurement of the forward production 
cross-section for the same target
and beam energy, relevant for the calculation of the 
far--to--near ratio of the K2K experiment. 
\vspace{1pc}
\end{abstract}

\maketitle

\section{Introduction}

The HARP experiment\cite{harp} was designed to perform a systematic and
precise study of hadron production for beam momenta between 1.5 and
$15~GeV/c$ and target nuclei ranging from hydrogen to lead. 
The detector was located at CERN, in the PS beam.The DAQ 
recorded 420 million events  
during the years 2001 and 2002.

The physics program of HARP includes: 
a) the measurement of pion yields for a variety of 
energies and targets relevant for he design of the proton driver of a
future neutrino factory\cite{nufact}; 
b) the measurement of pion yields on low $Z$ targets 
as well as on cryogenic oxygen and helium targets, 
useful to improve the precision of atmospheric neutrino flux
calculations\cite{barr}; and c) the measurement of 
pion and kaon yields, relevant for the calculation
of the neutrino fluxes of experiments such as  MiniBooNE\cite{miniboone} 
and K2K\cite{k2k}.

\begin{figure}[htbp]
\begin{center}
\hspace{0mm} \epsfig{file=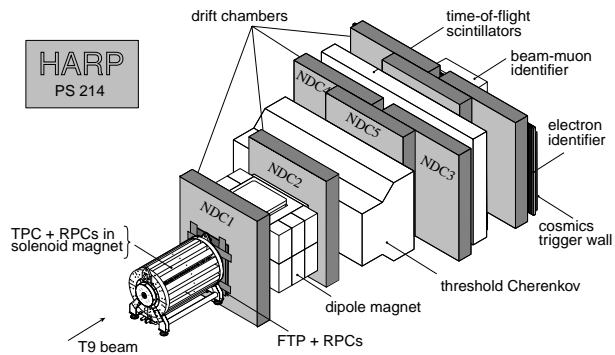,width=8cm}
\end{center}
\caption{Schematic layout of the HARP spectrometer.}
\label{fig:harp}
\end{figure}

HARP (Fig. \ref{fig:harp}) is
a large acceptance spectrometer, with two distinct regions. In the forward
part of the apparatus (up to polar angles of about 250 $~mrad$), 
the main tracking devices are a set of large drift chambers. 
Magnetic analysis is provided by a 0.4 $~T$ dipole magnet and  
particle identification relies in the combination 
of a threshold Cherenkov detector, a time-of-flight
wall and an electromagnetic calorimeter. In the rest of the solid angle
the main tracking device is a TPC, which is complemented by a set of RPC
detectors for time-of-flight measurements. The target is located
inside the TPC. 
In addition, sophisticated beam instrumentation 
(including three timing detectors and threshold Cherenkov detectors) 
provides identification of the incoming particle and allows
the interaction time at the target to be measured.
 
Given the immediate interest of the MiniBooNE and K2K experiments 
in a measurement of the production cross sections 
for pions and kaons at the energies and
targets relevant for their beam setups, 
the HARP collaboration has given priority to the
analysis of those particular data sets. In particular, we present in 
this article an initial result
relevant for the K2K experiment. Specifically, we 
measure pion yields from proton-nucleus 
collisions, for a thin Al target ( $5\% \lambda$ ) and
a beam energy of 12.9 $~GeV$. The K2K neutrino beam is 
produced from the decay of hadrons
(primarily pions) emanating from a  $2 \lambda $ Al target, 
at the KEK PS proton driver energies of 12.9 GeV. Further details about
the motivation of this analysis can be found in Ref.~\cite{anselmo}.

\section{Forward Tracking and Particle Identification} \label{sec:tracking}

\subsection{Tracking}

Tracking of forward-going particles is done by a set of large 
drift chambers(NDC) placed upstream and downstream of the dipole magnet.
The chambers were recuperated from the NOMAD
experiment and their properties have been described
elsewhere\cite{NOMAD_NIM_DC}. 
Each NDC module is made of four chambers, 
and each chamber of three  planes of wires with 
tilted angles $-5^o$, $0^o$ and $5^o$. 
The single-wire efficiency  is
of the order of $80\%$, and the spatial resolution 
approximately $340 \ \mu m$. The performance
of the NDC is inferior to the one measured in NOMAD, 
where the hit efficiency was close
to $~95\%$ and the resolution approached $150 \ \mu m$. 
This is mainly due to the 
fact that, for HARP, the NDC used a gas mixture and 
voltage settings different from the ones 
in NOMAD\footnote{$Ar(90\%)-CO_2(9\%)-CH_4(1\%)$, 
a non-flamable gas mixture.}.

Due to the low hit efficiency, the algorithm used in NOMAD, 
based in building up space points
(with an efficiency of about $.95^3 = 85 \% $) 
is not viable (for Harp we would obtain
$.80^3 = 51 \% $). Instead, the reconstruction 
builds $2D$ and $3D$  track segments 
in each NDC module (12 hits maximum), which are 
fitted to a straight line model via a
Kalman Filter fit\cite{Kalman}.

Next, the algorithm attempts all possible 
combinations (which include at least a $3D$  segment) 
to connect tracking objects in the
modules downstream of the dipole magnet. 
Combinations such as $3D$ + $2D$ or even
$3D$ + hits--not--associated are valid ways to 
build longer, $3D$ segments. 

To measure the momentum it is necessary to connect a ($3D$) 
segment downstream of the dipole with
at least one space point upstream of the dipole. 
Since one can impose the constraint that all tracks 
emanate from the event vertex\footnote{for a 
thin target the vertex is known with a precision of about $1 ~cm$
In the case of thick target, a vertex algorithm will provide
the event vertex to a comparable precision.}this
point is always known and therefore the necessary and sufficient 
condition for a particle emanating from the target to have its 
momentum measured is that a $3D$ segment can be measured by 
the combination of downstream NDC chambers.

A measurement in the first NDC module is not strictly necessary to 
analyse the track momentum. Ideally, one would like, of course, to
connect a $3D$ segment downstream with a $3D$ 
segment in the NDC1 {\em and} the vertex point. 
However, the tracking efficiency of the upstream 
module is sizeable lower than the efficiency of the 
downstream modules. This is due to, a) a higher 
hit density in NDC1\footnote{the hit density decreases
quadratically with the distance to the vertex, and 
therefore the effect is only significant for NDC1.} and b) 
the lack of redundance due to the fact that there is 
only a single module upstream of the dipole. A too restrictive 
condition to accept a track such as matching a $3D$
segment upstream of the dipole with a $3D$ segment 
downstream results not only in lower efficiency, 
but also in a systematic error, 
arising from the uncertainties in the hit density expected 
in NDC1 which would be
hard to estimate (hit density depends on
factors such as track multiplicity and opening angle,
which are model-dependent).
Instead, by building a track {\em always} when a
$3D$ segment is present downstream of the dipole, one is 
practically independent of 
the tracking efficiency in NDC1.

In order to quantify these considerations, 
we have considered three different types of tracks, 
depending on the matching between the downstream and 
upstream modules. Type I is a $3D$-$3D$ matching. 
Type II matches a $3D$ (downstream) with a $2D$ (upstream). 
Type III  are those where one has
been unable to find either a $3D$ or a $2D$ segment 
in NDC1 and thus the track is built using
the downstream modules and the vertex constraint.

\subsection{Particle Identification}

Particle identification (PID) in the forward region of 
the spectrometer combines the information 
provided by beam detectors and three systems located 
downstream the dipole magnet. Namely,  a threshold 
Cherenkov detector (CKOV), a time--of--flight wall (TOF), and an
electromagnetic calorimeter (electron identifier, EID). 
These subsystems have been described in Ref. ~\cite{anselmo}.
Its combined information results in good PID over the whole
range of relevant momenta, as well as redundancy due to overlaps. Pion/proton 
separation is provided by TOF up 
to $4.5~GeV/c$, and by the CKOV above $3~GeV/c$. 
Electron/pion separation is covered by the CKOV below $3~GeV/c$ and by the EID 
above $2~GeV/c$. Finally the kaon contamination can  be estimated with the  
CKOV above $3~GeV/c$ and with the (TOF) below this energy.

\section{The analysis} \label{sec:analysis}

About one sixth (1 million events) of
 the ``K2K thin target'' data has been analysed. 
The unnormalised pion production differential 
cross section can be computed as follows:

\begin{equation}
\sigma_i^\pi = \frac{1}{\varepsilon_i^{acc}} 
               \frac{1}{\varepsilon_i^{track}}
\sum_{t=1}^{3}  \left[  M_{ij}^{(t)} 
        \frac{1}{\varepsilon_j^{(t)-\pi}} 
        \eta_j^{(t)-\pi} \cdot
        N_j^{(t)-\pi} 
      \right]
\label{eq:cross}
\end{equation}
%
%
where the sum runs over track types (I = $3D$-$3D$, II= $3D$-$2D$, 
III=$3D$-vertex) and
the indices $i$ and $j$ correspond to true and 
reconstructed $(p,\theta)$ bins respectively. 
$\varepsilon_i^{acc}$ is the geometrical acceptance, 
$\varepsilon_i^{track}$ is the tracking efficiency, 
$M_{ij}$ is the migration matrix from reconstructed bin $j$ to true
bin $i$, $\varepsilon_j^{\pi}$ is the pion identification efficiency, 
$\eta_j^{\pi}$ is the pion purity and $N_j^{\pi}$ is the
observed pion yield. The pion purity is defined as 
$\eta_j^{\pi} = N_j^{true-\pi} / N_j^{\pi}  = 
(N_j^{\pi}-N_j^{bkg}) / N_j^{\pi}$, 
where $N_j^{true-\pi}$ is the number of true 
observed pions in the bin $j$, 
and $N_j^{bkg}$ is the number of particles 
misidentified as pions in the 
same bin $j$. That means that $\eta_j^{\pi}$  
is just the probability of 
correct identification of a pion. The pion 
identification efficiency, $\varepsilon_j^{\pi}$  
is the fraction of times that a tracked pion is identified as such. 

\begin{figure}[htbp]
  \begin{center}
  \epsfig{file=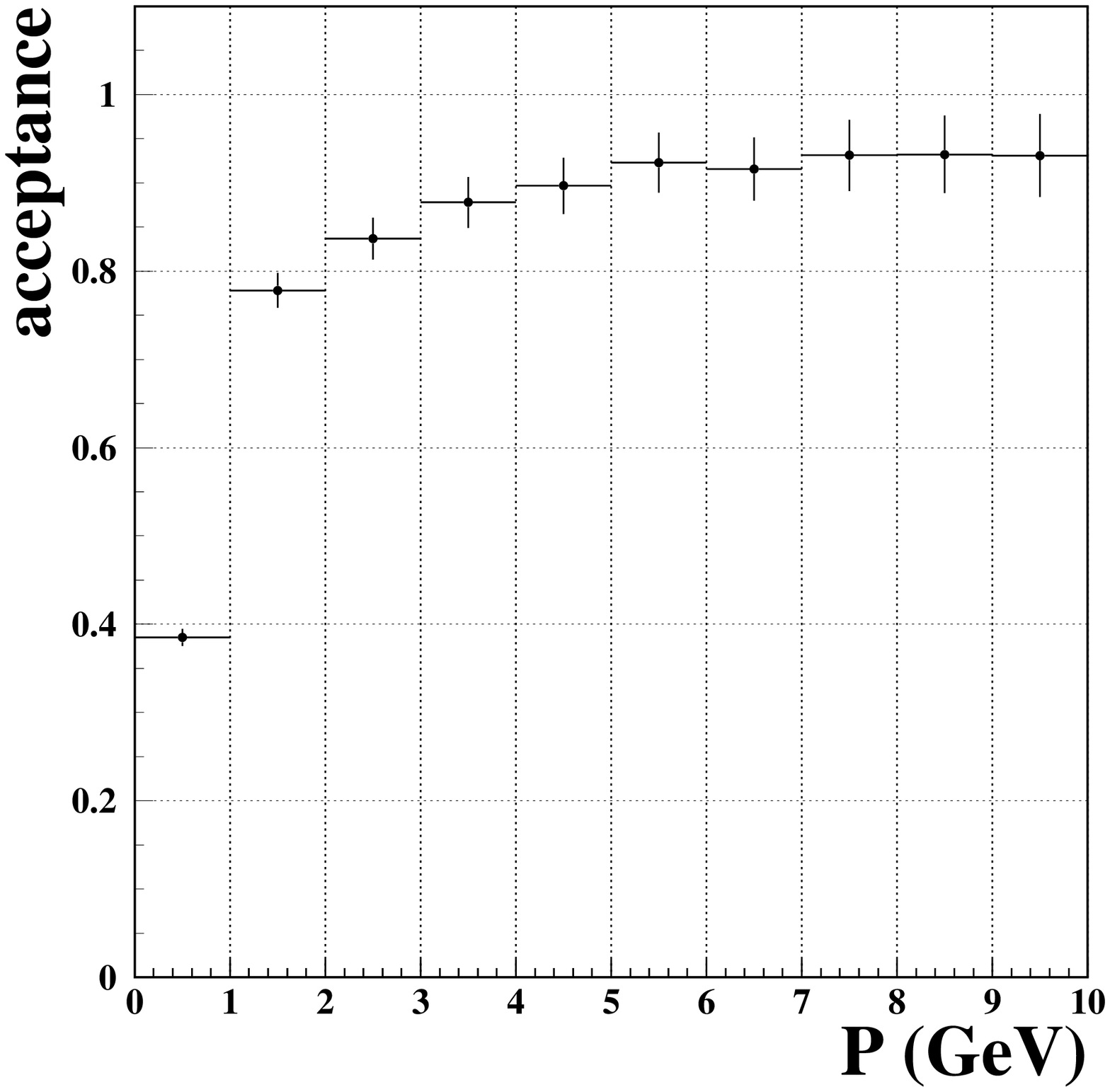,width=7cm}
  \epsfig{file=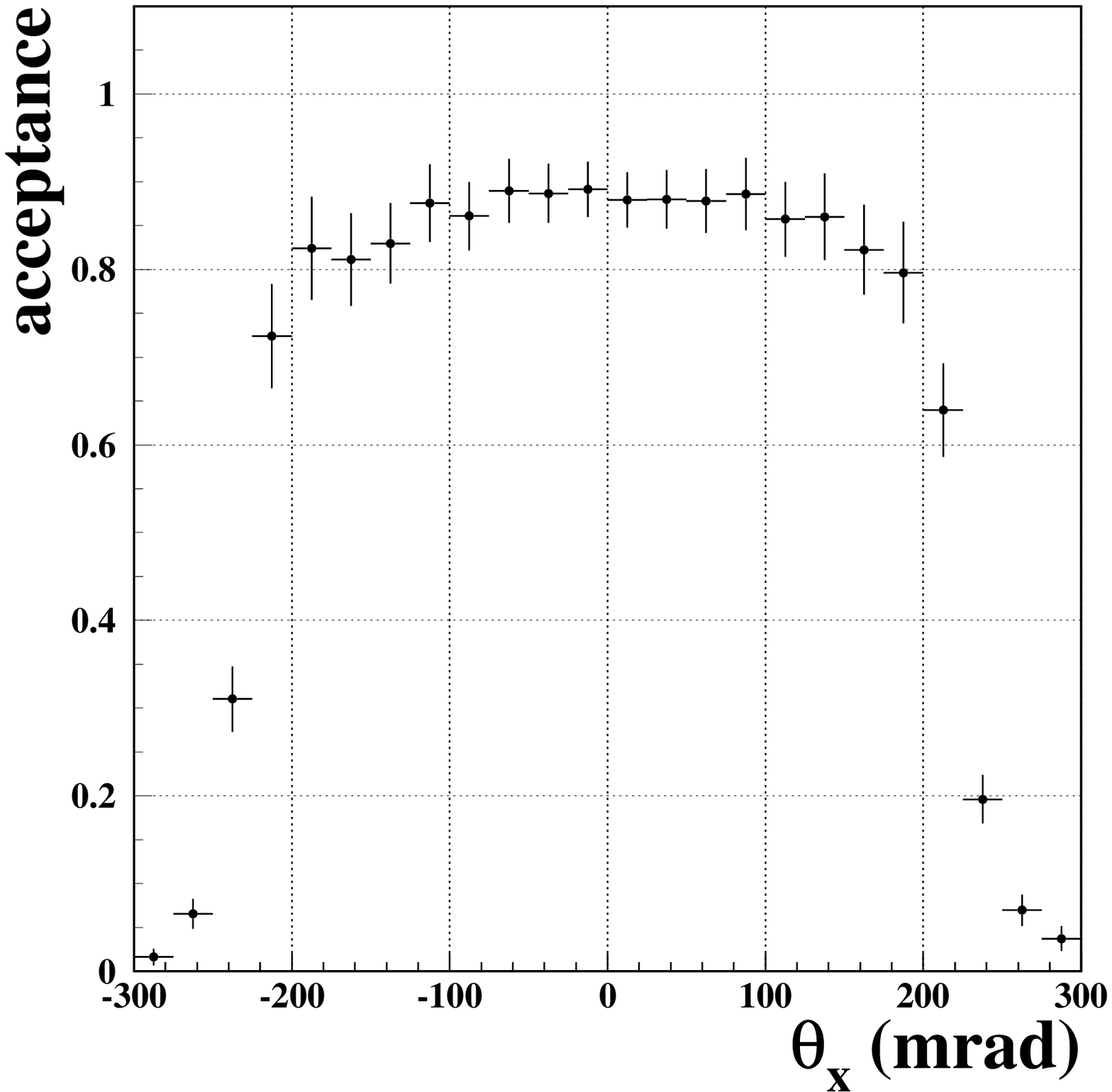,width=7cm}
  \caption{\label{fig:down_eff_p_th}
Acceptance in P for $|\theta_y|<50 mrad$,  $|\theta_x|<200 mrad$. 
Acceptance in $\theta_x$ for $|\theta_y|<50 mrad$ and $P>1 GeV$}
  \end{center}
\end{figure}

Fig.~\ref{fig:down_eff_p_th} shows the spectrometer acceptance. 
For clarity of ilustration, the
momentum acceptance has been computed in the region of good angular
acceptance, ($|\theta_y|<50 ~mrad$,  $|\theta_x|<200 ~mrad$) 
and conversely, the geometrical acceptance has been computed
in the region of good momentum acceptance ($P>1 GeV$). As it can
be seen the acceptance is good for pion momenta above 1 $~GeV/c$, and still 
acceptable above $0.5 ~GeV$. As illustrated in Ref.~\cite{anselmo} 
this permits
the measurement of K2K neutrino fluxes up to $~0.25~GeV$, which amply
covers the energy region where the atmospheric oscillation 
affects the neutrino spectrum maximally.
The angular acceptance is also very well
matched to K2K physics requirement.
 
\begin{figure}[htbp]
  \begin{center}
  \epsfig{file=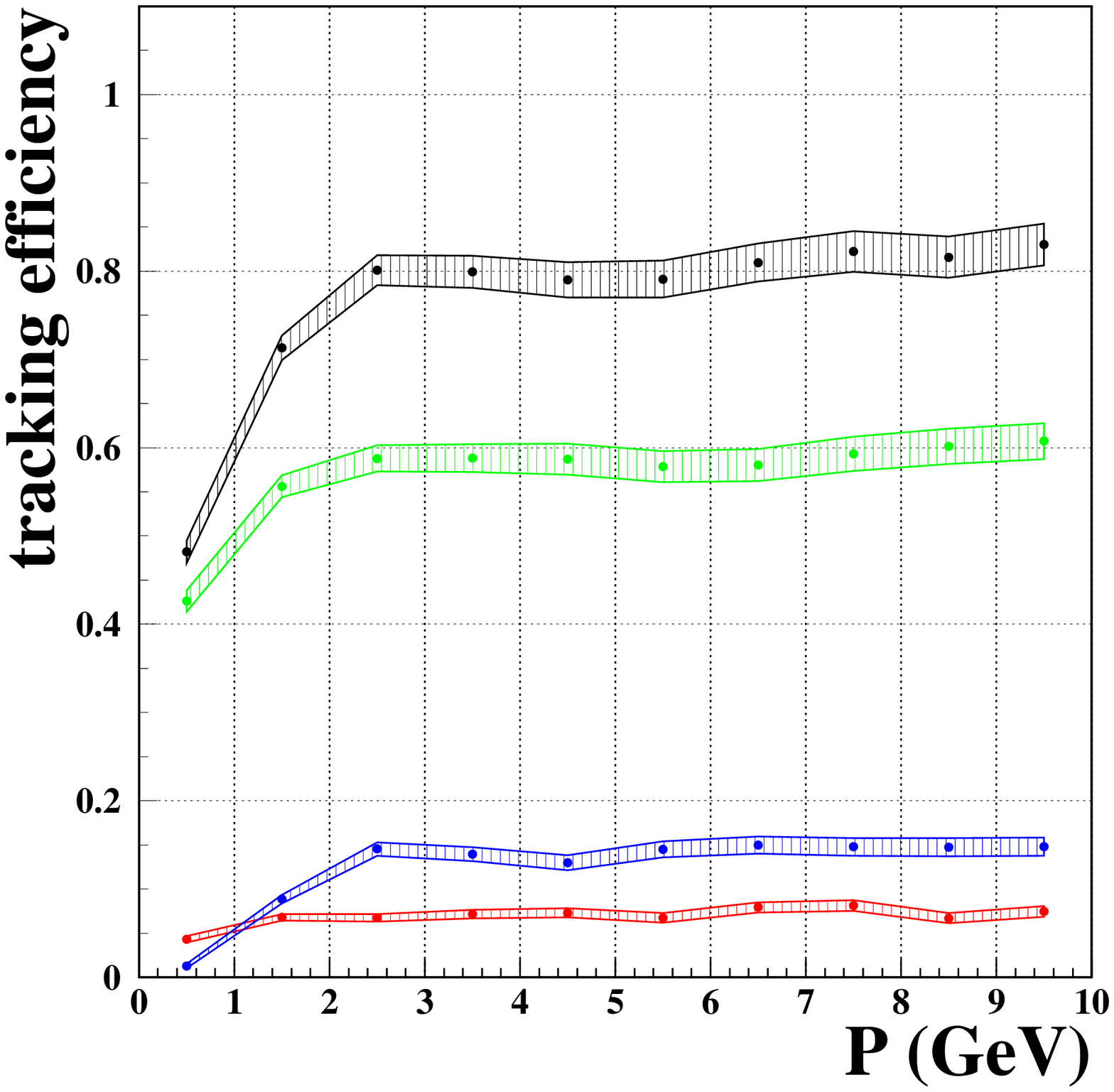,width=7cm}
  \epsfig{file=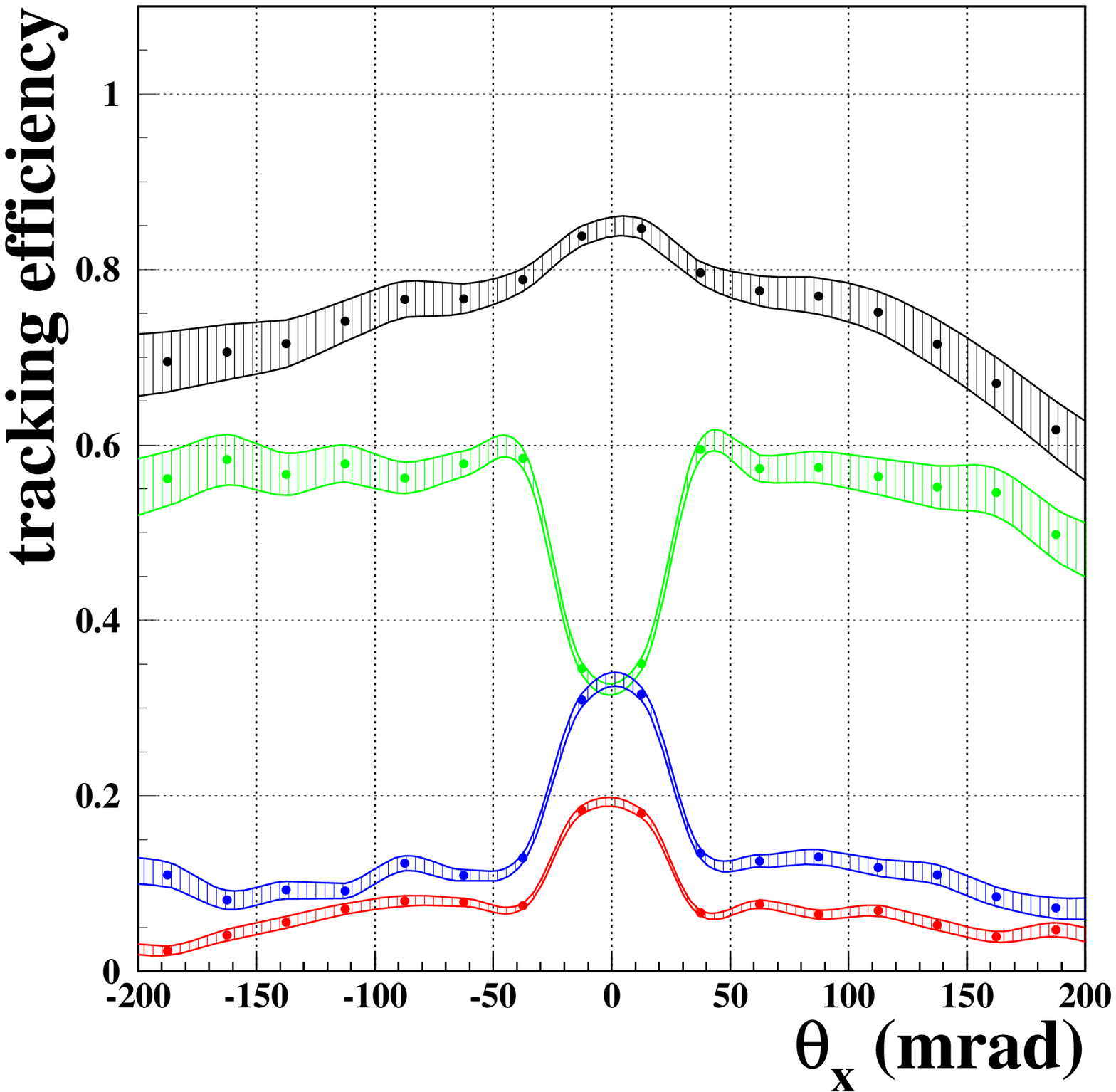,width=7cm}
  \caption{\label{fig:track_eff_mc1_normalized_p_th}
 Tracking efficiency as a function of 
$p$ (left) and $\theta_x$ (right).
  The colors indicate: black=total, green=type I, red=type II,
  blue= type III.}   
  \end{center}
\end{figure}

The tracking efficiency is shown in 
Fig.~\ref{fig:track_eff_mc1_normalized_p_th} 
as a function of $p$ and $\theta_x$. The contributions of 
track types I,II and III are normalized individually, so that the total
efficiency is the sum of the three. Notice that the loss in efficiency at
low polar angles for tracks of type I (due to saturation effects in NDC1)
is largely compensated by tracks 
of type II and III, so that the total tracking efficiency is rather flat.

\begin{figure}[htbp]
  \begin{center}
  \epsfig{file=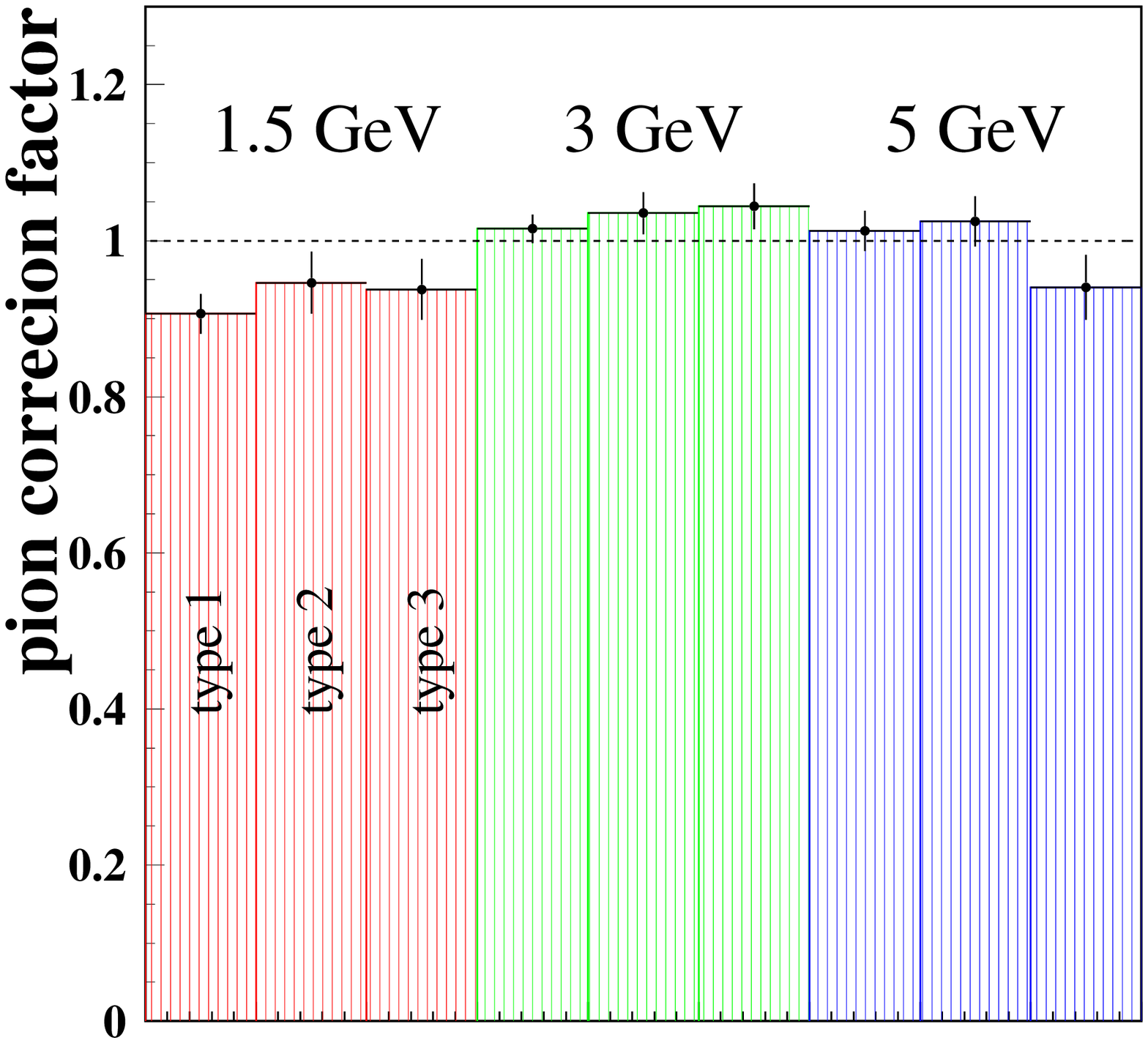,width=7cm, height=5cm}
  \caption{\label{fig:pion_corr_factor}
Pion correction factor. 
  The colors indicate: red=1.5GeV, green=3GeV, blue=5GeV}   
  \end{center}
\end{figure}

Finally, the yield must be corrected by the pion efficiency and purity, 
which is computed from the data themselves, using samples
of pions, protons electrons and kaons of various energies tagged with the
help of the beam detectors. The pion correction factor defined as the
ratio of purity over efficiency
is shown in  Fig. \ref{fig:pion_corr_factor}, for the
three track types and for 1.5, 3 and 5 GeV beam particles.

Preliminary studies show an almost diagonal migration matrix, where
migration is mainly due to finite momentum and angular resolutions, 
but not to systematic effects (momentum and angular biases, etc).

\begin{figure}[htbp]
  \begin{center}
  \epsfig{file=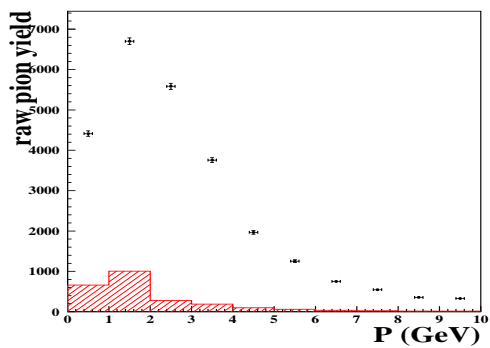,width=7cm, height=5cm}
  \epsfig{file=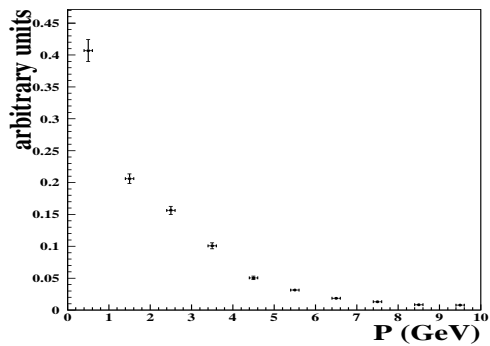,width=7cm, height=5cm}
  \caption{\label{fig:cross}
Upper; Raw pion yield. Lower; Yield corrected by
pion correction factor, efficiency and acceptance.
The red filled histo corresponds to the proton+electron background}   
  \end{center}
\end{figure}

Figure ~\ref{fig:cross} shows the pion yields 
as a function of momentum after succesive
corrections (e.g., acceptance, tracking efficiency and pion correction
factor). No attempt to deconvolute the data (e.g, correct for migrations)
have been done yet, and no absolute normalization is computed. 

\section{Conclusions} \label{sec:conclu}

We have presented a preliminary analysis of one of HARP 
data samples, corresponding to a thin Al target, at beam energies 
of $ 12.9~GeV$. This analysis is relevant for the K2K experiment and a 
first step towards an upcoming measurement of the fully corrected
production cross section.

\end{document}